\documentclass[preprint,onecolumn,nofootinbib]{revtex4}

\usepackage{float}
\usepackage{graphicx}
\usepackage{array}
\usepackage{delarray}
\usepackage{amsmath}
\usepackage{amssymb}

\begin{document}

\newcommand{\myvec}[1]{\accentset{\rightharpoonup}{#1}}
\newcommand{\ket}[1]{| #1 \rangle}
\newcommand{\bra}[1]{\langle #1 |}
\newcommand{\op}[1]{{\mathbf #1}}
\newcommand{\ops}[1]{{\boldsymbol #1}}
\newcommand{\Mit}{\mathrm}
\newcommand{\Tr}[2][]{\mathrm{Tr}_{#1} \! \left[ #2 \right]}

\newcommand{\be}{${}^9\mbox{Be}^+$ \:}
\newcommand{\Sstate}{{}^2{\rm S}_{1/2} \:}
\newcommand{\Pone}{{}^2{\rm P}_{1/2} \:}
\newcommand{\Pthree}{{}^2{\rm P}_{3/2} \:}
\newcommand{\Plev}{\mathrm{P}}
\newcommand{\Fd}{F=2, m_F=-2}
\newcommand{\Fu}{F=1, m_F=-1}
\newcommand{\Fs}{{{\mathcal F}_\mathrm{S}}}
\newcommand{\Fone}{{{\mathcal F}_{1/2}}}
\newcommand{\Fthree}{{{\mathcal F}_{3/2}}}
\newcommand{\Fp}{{{\mathcal F}_\mathrm{P}}}
\newcommand{\Fk}{{{\mathcal F}_k}}
\newcommand{\Eps}{{\hat{\epsilon}}}
\newcommand{\phif}{\phi_\Mit{fluct}}

\newcommand{\ua}{{\uparrow}}
\newcommand{\da}{{\downarrow}}
\newcommand{\uan}{{\uparrow^{\mathrm{N}}}}
\newcommand{\dan}{{\downarrow^{\mathrm{N}}}}
\newcommand{\hc}{\mbox{h.c.}}
\newcommand{\cc}{\mbox{c.c.}}
\newcommand{\deltak}{\myvec{\Delta k}}
\newcommand{\omegaud}{\omega_{\da\ua}}
\newcommand{\omegafs}{\omega_{\rm FS}}
\newcommand{\sumj}{\sum_{j=1,2}}
\newcommand{\sumk}{\sum_{k=\{ 1/2,3/2 \}}}
\newcommand{\Omegaud}{\Omega_{\da\ua}}
\newcommand{\deltaud}{\delta_{\da\ua}}
\newcommand{\Oop}{\boldsymbol{\mathcal{O}}}
\newcommand{\nbar}{\overline{n}_\Mit{COM}}
\newcommand{\heat}{\Gamma_\Mit{heat}}

\newcommand{\mydash}{~$\leftrightarrow$~}

\newcommand{\antih}{$\overline{\mbox{H}}$ }
\newcommand{\htwo}{$\mbox{H}_2$}
\newcommand{\yb}{$\mbox{Yb}^+\:$}

\newcommand{\widtha}{14cm}
\setlength{\abovedisplayskip}{1mm} 

\title{Laser cooling of new atomic and molecular species with ultrafast pulses}

\author{D. Kielpinski}

\affiliation{Research Laboratory of Electronics and Center for Ultracold Atoms, Massachusetts Institute of Technology,
Cambridge MA 02139}

\date{\today}

\begin{abstract}

We propose a new laser cooling method for atomic species whose level structure makes traditional laser cooling difficult. For instance, laser cooling of hydrogen requires single-frequency vacuum-ultraviolet light, while multielectron atoms need single-frequency light at many widely separated frequencies. These restrictions can be eased by laser cooling on two-photon transitions with ultrafast pulse trains. Laser cooling of hydrogen, antihydrogen, and many other species appears feasible, and extension of the technique to molecules may be possible.

\end{abstract}

\maketitle

Laser cooling and trapping are central to modern atomic physics. The low temperatures and long trapping times now routinely achieved by these means have led to great advances in precision spectroscopy and cold collision studies. These conditions also provide a suitable starting point for evaporative cooling to Bose-Einstein condensation. However, traditional laser cooling methods are easily applied only to atomic species that exhibit strong, closed transitions at wavelengths accessible by current laser technology. Only $\sim 20$ species have been laser-cooled, mostly alkali and alkali-earth metals and the metastable states of noble gases \cite{halbook}.\\

Two obstacles impede the further extension of laser cooling techniques. First, the lowest energy transitions of many atoms of interest, including hydrogen and carbon, lie in the vacuum ultraviolet (VUV). Not enough laser power is available in this spectral region to drive effective laser cooling. Second, the complex level structure of many atoms (and all molecules) permits decay of an excited electron into a number of metastable levels widely separated in energy. Each metastable decay channel must typically be repumped by a separate laser, so the laser system becomes unwieldy.\\

Laser cooling of hydrogen (H), deuterium (D), and antihydrogen (\antih) has remained elusive owing to the first obstacle, the lack of power available at the required 121 nm VUV wavelength. Improved spectroscopy of the 1S -- 2S two-photon transition at 243 nm is the most obvious payoff for laser cooling these atoms. The 1S -- 2S transition plays a unique role in metrology. Measurements of its frequency in H are accurate at the $10^{-14}$ level \cite{Niering-Haensch-absolute-H-1S2S} and assist in determining the value of the Rydberg constant \cite{Schwob-Biraben-H-rydberg-constant}. The isotope shift of the 1S -- 2S transition between H and D gives the most accurately determined value of the D nuclear radius, tightly constraining nuclear structure calculations \cite{Huber-Haensch-H-isotope-shift}. Possibly the most exciting application is a comparison between H and \antih 1S -- 2S frequencies, using the low-energy \antih recently produced at CERN \cite{ATHENA-antiH-production,ATRAP-antiH-production}. Such comparisons can test CPT symmetry to unprecedented accuracy, probing physics beyond the Standard Model \cite{hbarcpt1,hbarcpt2}. The H 1S -- 2S measurement is currently limited by the $\sim 6$ K temperature of the H beam and could be improved by two orders of magnitude with colder atoms \cite{Niering-Haensch-absolute-H-1S2S}, e.g. in an atomic fountain \cite{hfount}. The \antih formation temperature in the CERN experiments is likely to be of the same order, limiting the corresponding \antih measurement.\\

Cooling of H below a few K currently requires direct contact with superfluid helium \cite{Hess-Kleppner-H-magnetic-trap,vanRoijen-Walraven-H-magnetic-trap}. This method appears unlikely to work for \antih. Attempts to cool D in this way have been unsuccessful because of the high binding energy of D on liquid helium \cite{deCarvalho-Greytak-buffer-gas-H}. Even for H it is cumbersome, requiring a dilution refrigerator and a superconducting magnetic trap, which severely restricts optical access. Current proposals for laser cooling H, D, and \antih involve generation of Lyman $\alpha$ (121 nm) light for excitation of the 1S -- 2P transition. The small amount of light available means that cooling is extremely slow, on the timescale of minutes in the only experiment reported so far \cite{Setija-Walraven-H-laser-cooling}.\\

Many atomic species of chemical and biological interest, including carbon, nitrogen, and oxygen, suffer from the second obstacle. These species have several valence electrons, and are difficult to laser-cool because of the many widely separated frequencies required for repumping atomic dark states. On the other hand, spectroscopy on ultracold samples of these atoms would greatly improve understanding of their long-range interactions and chemical bonding behavior, similar to studies already performed for most alkalis (see \cite{stwparev} for a recent review). Since these atoms display rich interactions and are common building blocks of everyday objects, this kind of information can potentially impact many fields, from biology to astrophysics. Simultaneous cooling of H and C could even lead to synthesis of organic molecules at ultracold temperatures, as in current experiments that produce ultracold molecules from laser-cooled alkali gases \cite{Kerman-DeMille-RbCs-production,Wang-Stwalley-KRb-production}.\\

We propose a laser cooling scheme that uses ultrafast pulse trains to address both obstacles, opening many new atomic species to laser cooling. The ultrafast pulse trains from mode-locked lasers exhibit high spectral resolution \cite{tedspec1,tedspec2,Snadden-Ferguson-MLL-two-photon-Rb}. The high peak powers of ultrafast pulses enable efficient nonlinear optics far into the UV, greatly increasing the time-averaged optical power available at short wavelengths \cite{visdbl,persaud}. At the same time, the many frequencies generated in short pulses can perform the function of repumping lasers, reducing the complexity of laser systems for cooling atoms with multiple valence electrons. Because of their high peak powers and high spectral resolution, ultrafast pulse trains are much more effective than single-frequency lasers for two-photon laser cooling. We demonstrate the usefulness of our scheme for laser cooling H and \antih in currently used magnetic traps, and discuss a potential cooling scenario for atomic carbon. An extension of the scheme for laser cooling of molecules appears promising.\\

Laser cooling requires velocity-selective scattering to compress the atomic velocity distribution. A pulse train from a mode-locked laser can have high spectral resolution, sufficient to resolve atomic transitions at their natural linewidth \cite{tedspec1,tedspec2,Snadden-Ferguson-MLL-two-photon-Rb}. As shown in Fig. 1, the spectrum of such a pulse train is a comb of sharp lines with frequencies $\nu_k = \nu_\Mit{car} + k \nu_\Mit{rep}$, where $k$ is an integer, $\nu_\Mit{car}$ is the optical carrier frequency, and $\nu_\Mit{rep}$ is the pulse repetition rate. If one comb line is nearly resonant with a Doppler-broadened atomic transition of width $\Gamma_D$, and $\nu_\Mit{rep} \gg \Gamma_D$, the scattering rates induced by the neighboring comb lines are reduced by a factor $(\Gamma_D/\nu_\Mit{rep})^2$. The rapid falloff of scattering rate with detuning ensures that, although there are many comb lines, the dominant contribution to the total scattering rate comes from the single near-resonant comb line. Hence velocity-selective scattering proceeds as for a CW laser and the Doppler cooling limit is $h \Gamma/2$, where $\Gamma$ is the natural width of the atomic transition. Velocity-selective scattering can also occur for $\nu_\Mit{rep} < \Gamma$, if the pulses are detuned by more than their bandwidth $\Delta\nu$ from atomic resonance, but the Doppler cooling limit is then $h \Delta\nu/2$, corresponding to temperatures of a few K for pulse durations of a few ps \cite{Blinov-Monroe-pulsed-cooling-preprint}. The related ``white-light" cooling schemes use an additional CW source near the atomic resonance to achieve temperatures $\ll h \Delta\nu/2$ \cite{Hoffnagle-white-light-cooling,whiteexpt1,Zhu-Hall-white-light-cooling}, but this is a difficult requirement in the cases we will consider.\\

\begin{figure}
\begin{center}
\includegraphics*[width=\widtha]{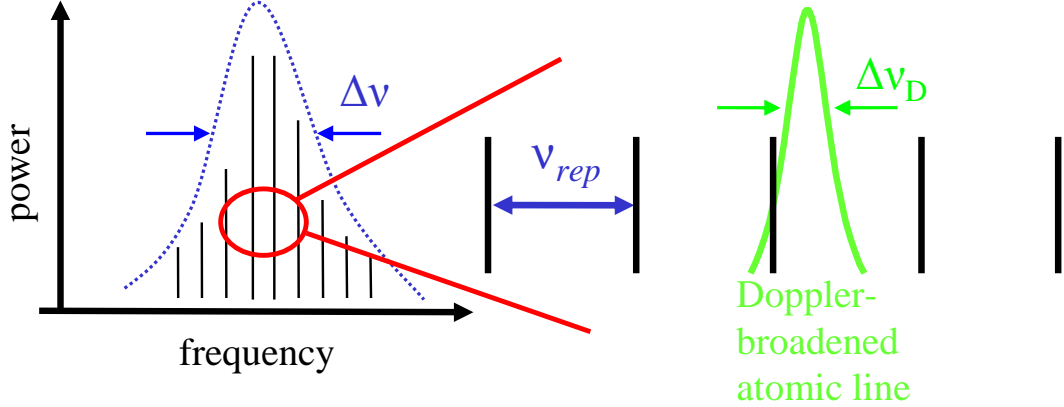}
\caption{Velocity-selective scattering with a high-repetition-rate pulse train. Left: Optical spectrum generated by a mode-locked pulse train, consisting of equally spaced sharp lines (black) with a spectral envelope of bandwidth $\Delta\nu$ (blue). Right: Velocity selective scattering occurs when one laser line is nearly resonant with a Doppler-broadened atomic transition of width $\Delta\nu_D$. For repetition rate $\nu_\Mit{rep} \ll \Delta\nu_D$, all other laser lines induce negligible scattering.}
\end{center}
\end{figure}

In most laser-cooling schemes, the efficiency of laser cooling depends critically on the scattering rate, since a scattering event changes the atomic momentum, on average, by one photon recoil. Fig. 2 compares single-photon and two-photon scattering for mode-locked and CW excitation. For $\nu_\Mit{rep} \gg \Gamma$, the scattering rate on a single-photon transition is seen to be a factor of $\nu_\Mit{rep}/\Delta\nu$ smaller for mode-locked than for CW excitation, given equal average laser intensity. Since a given laser can achieve approximately the same time-averaged power whether it is operated CW or mode-locked, mode-locked excitation is less efficient than CW excitation for single-photon scattering. However, mode-locked and CW excitation can be equally efficient for two-photon scattering. A train of mutually coherent transform-limited pulses with time-averaged intensity $\overline{I}_\Mit{ML}$ induces a two-photon scattering rate $S^{(2)}(\overline{I}_\Mit{ML})$ approximately equal to the rate $S^{(2)}(I_\Mit{CW})$ induced by a CW laser of the same intensity \cite{Baklanov-Chebotaev-pulsed-two-photon}. Roughly speaking, each pair of mode-locked comb lines induces a transition path, and all pathways add coherently for transform-limited pulses. If the total average power is divided equally among all comb lines, the transition rate becomes independent of the number of comb lines. In the ultraviolet, mode-locked laser systems offer considerably higher average powers than CW laser systems, so two-photon cooling rates can increase by orders of magnitude over their CW values. This advantage makes two-photon mode-locked cooling competitive with single-photon CW cooling in the cases studied below.\\

\begin{figure}
\begin{center}
\includegraphics*[width=8cm]{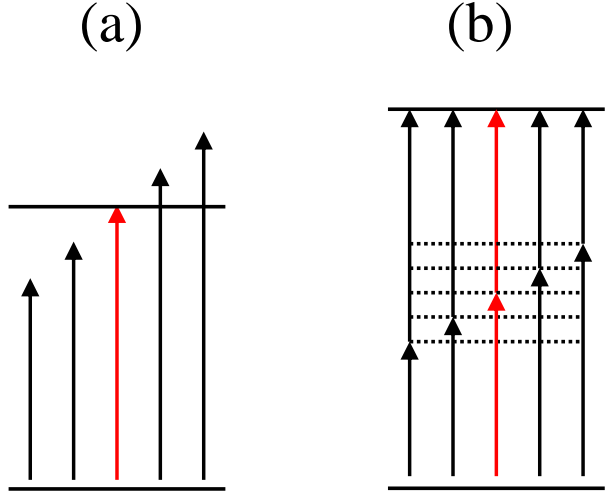}
\caption{Comparison of mode-locked and CW excitation of (a) single-photon and (b) two-photon transitions using energy-level diagrams. In (a), the frequency comb of the mode-locked laser (black) has only one component resonant with the atomic transition, while all the CW light (red) is resonant. In (b), we show a two-photon transition for which all intermediate states are far from single-photon resonance. The mode-locked laser induces many transition paths whose amplitudes add coherently, while the CW light follows only one path. The two-photon transition rates turn out to be roughly equal for equal average power.}
\end{center}
\end{figure}

The high spectral resolution of ultrafast pulse trains and their efficient excitation of two-photon transitions suggest that one can use mode-locked lasers to perform laser cooling on two-photon transitions when single-photon cooling is not possible. For the species H, C, O, N, F, and Cl, the lowest-energy single-photon transitions are all blue of 170 nm, precluding single-photon cooling, but these species also all exhibit two-photon transitions red of 170 nm. Single-photon cooling is relatively ineffective for these species because the available CW power is insufficient. Continuous-wave light with MHz bandwidth at $\lesssim 170$ nm has only been generated by four-wave mixing in atomic vapor \cite{Eikema-Haensch-Lya-H-excitation}. This method is highly technically challenging and yields only tens of nW of radiation. Two-photon scattering of CW light is relatively weak, since the available power is usually tens of mW. On the other hand, frequency conversion of ultrafast pulses can reach near-unit efficiency from infrared to visible \cite{visdbl} and from visible to UV \cite{persaud}, so average powers of $\sim 1$ W should be achievable for wavelengths $\gtrsim 170$ nm. Using ultrafast pulses increases the two-photon scattering rate by a factor $\sim 10^4$ over the CW case, simply owing to the higher nonlinear conversion efficiency.\\

In particular, mode-locked laser cooling on the 1S -- 2S two-photon transition at 243 nm is a good prospect for cooling magnetically trapped H, D, and \antih to Doppler-limited temperatures of a few mK. A possible excitation scheme is shown in Fig. 3. While the 2S state is metastable, one can quench the 2S state to the $2\mbox{P}_{3/2}$ state using microwave radiation near 10 GHz. If the two-photon laser is tuned to the $\ket{F,m_F} = \ket{1,1}$ -- $\ket{1,1}$ component of the 1S -- 2S transition and the quenching radiation is $\sigma^+$-polarized, the atoms are optically pumped to a stretched state and can remain magnetically trapped under laser excitation. The upper limit to the usable two-photon intensity comes from one-photon ionization of the excited state by 243 nm light. The photoionization rate from a given initial state is the same for mode-locked and CW excitation; since the final state is a continuum with slowly varying matrix element, all comb lines contribute equally. The photoionization rate from the excited state is then $R_\Mit{PI} = 11.4 \: \overline{I}_\Mit{ML} \: \mbox{Hz}\: \mbox{W}^{-1} \: \mbox{cm}^2$ \cite{sandberg}. If an atom undergoes $N_\gamma$ scattering events in cooling, we require $R_\Mit{PI}/\Gamma_\Mit{PI} \ll N_\gamma$ to avoid photoionization, so the maximum quenching is generally desirable. When the quenching radiation strongly saturates the $2\mbox{S}_{1/2}$ -- $2\mbox{P}_{3/2}$ transition, $\Gamma = 50$ MHz and the two-photon scattering rate at resonance is $R_2 = 2.8 \times 10^{-7} \: \overline{I}_\Mit{ML}^2 \: \mbox{Hz}\: \mbox{W}^{-2}\: \mbox{cm}^4$ \cite{sandberg}.\\

\begin{figure}
\begin{center}
\includegraphics*[width=\widtha]{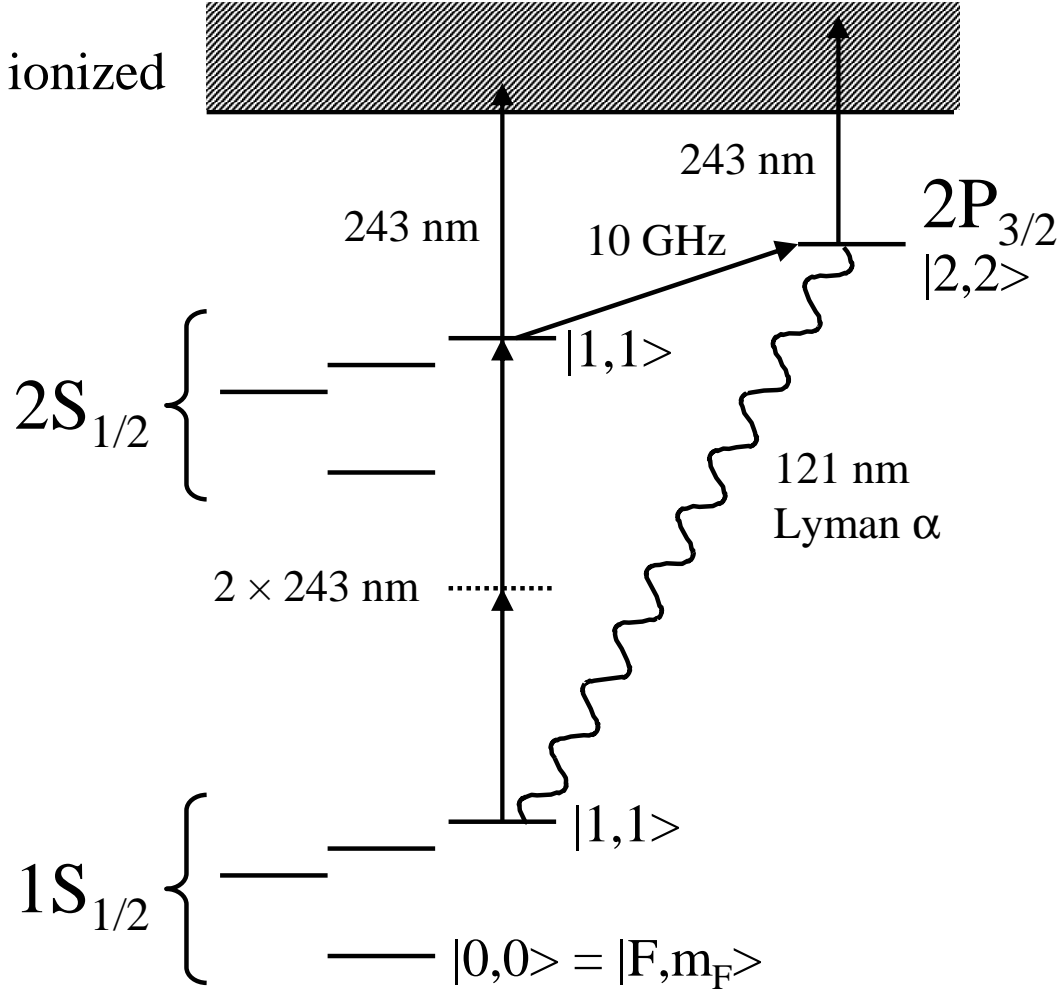}
\caption{Excitation scheme for laser cooling of magnetically trapped H or \antih. The 243 nm light excites the atoms from the magnetically trapped $1\mbox{S}_{1/2} \:\ket{F,m_F} = \ket{1,1}$ state to the $2\mbox{S}_{1/2} \:\ket{1,1}$ state. Radiation near 10 GHz quenches the metastable 2S state to the $2\mbox{P}_{3/2} \:\ket{2,2}$ state. The atoms reradiate on the $1\mbox{S}_{1/2} \ket{1,1}$ -- $2\mbox{P}_{3/2} \ket{2,2}$ transition at 121 nm, returning to the magnetically trapped state. While in the $2\mbox{S}_{1/2}$ or $2\mbox{P}_{3/2}$ state, an atom can be photoionized by a single 243 nm photon. For clarity, only the relevant $2\mbox{P}_{3/2}$ substate is shown.}
\end{center}
\end{figure}

Experiments on trapped hydrogen and antihydrogen would benefit from our proposed laser cooling technique. Proposed magnetic traps for antihydrogen \cite{Walraven-antiH-trap,Squires-Gabrielse-penning-ioffe-trap,Holzscheiter-Charlton-cold-antiH-rev} are similar to those currently used for hydrogen, so we estimate typical laser cooling parameters for both cases by considering the H trap apparatus used at MIT \cite{Hess-Kleppner-H-magnetic-trap}. In that experiment, cryogenically cooled H is loaded into a Ioffe-Pritchard magnetic trap, where up to $10^{13}$ H atoms equilibrate to a temperature of 40 mK (set by trap depth) with peak number density $2 \times 10^{13} \:\mbox{cm}^{-3}$ \cite{Masuhara-Kleppner-H-evaporative-cooling,Moss-Kleppner-thesis}. From the magnetic trap parameters and the loading temperature, we estimate the radius of the H sample as 2 mm and its length as 40 mm \cite{Hess-Kleppner-H-magnetic-trap,Cesar-Kleppner-trapped-H-1S2S}. A quenching radiation power of 1.6 W with diffraction-limited focusing is sufficient to achieve a 50 MHz Rabi frequency on the $2\mbox{S}_{1/2}$ -- $2\mbox{P}_{3/2}$ transition. At a two-photon cooling intensity of $60 \:\mbox{kW} \:\mbox{cm}^{-2}$, the resonant scattering rate is 1.0 kHz and the photoionization rate under resonant two-photon excitation is 7 Hz. The sample geometry only allows us to achieve this high intensity along the trap axis. The deceleration is $3.2 \times 10^4 \:\mbox{m} \:\mbox{s}^{-2}$, and an atom can generally be cooled to the one-dimensional Doppler limit in 8 ms if it stays in the cooling light. With these parameters, only 5\% of atoms will be lost to photoionization. The cooling time is much shorter than the axial period of the trap, indicating that a transversely-guided atomic beam could also be cooled by our technique.\\

The high UV powers from mode-locked pulse trains are essential to maintain such high intensities over a reasonable area. Such light can be generated by frequency-doubling a mode-locked Ti:S pulse train twice \cite{visdbl,persaud}, yielding average powers up to 1 W. Resonant enhancement cavities at 243 nm regularly achieve power buildup factors of 30 \cite{ferdi1,dana1}, so the waist radius of the cooling light can be $200 \:\mu$m. For the MIT magnetic trap parameters, the cooling light then overlaps $10^{-2}$ of the sample volume. As the sample cools, the spatial and spectral overlap with the cooling light improves, but disregarding these factors we obtain a one-dimensional Doppler cooling time for the whole sample of $\sim 20$ s. Cross-dimensional thermalization from atomic collisions should cool the entire sample to the 2.4 mK Doppler limit in $\sim 60$ s.\\

While this scheme is clearly less efficient than laser cooling of alkali atoms, it is competitive with other methods for laser-cooling H and \antih. Mode-locked two-photon cooling compares well to cooling on the 121 nm 1S -- 2P transition owing to the technical difficulties of generating and manipulating 121 nm light. The first 121 nm sources were developed over 20 years ago \cite{Mahon-Koopman-first-Lya-source} and laser cooling of H at 121 nm was first reported over 10 years ago \cite{Setija-Walraven-H-laser-cooling}, but the highest 121 nm power reported is still only 20 nW \cite{Eikema-Haensch-cw-Lya-source}. Current proposals for 121 nm laser cooling expect resonant scattering rates less than 1 kHz for a $200 \:\mu$m beam waist \cite{Walz-Haensch-cw-Lya-source,Eikema-Haensch-Lya-H-excitation}. Mode-locked two-photon cooling also improves on CW two-photon cooling. Only 20 mW of 243 nm CW light is available \cite{sandberg,ferdi1}, so the resonant scattering rate would drop to $\sim 1$ Hz for CW two-photon cooling over the same beam waist.\\

Our cooling scheme opens up further possibilities for laser cooling of atomic species with multiple valence electrons, which comprise most of the periodic table. These atoms often have many low-lying metastable states that are coupled by spontaneous emission during cooling. Efficient cooling requires repeated velocity-selective excitation of all transitions, so a narrowband radiation source must address each transition to avoid optical pumping into a dark state. While this task requires many CW lasers, a single mode-locked laser is sufficient. The octave-spanning laser oscillators currently available \cite{bartoctave} can easily cover the entire spectral range needed for excitation of all transitions. Although the transitions are spaced more or less randomly with respect to the comb of frequencies generated by the pulse train, the gaps between transition and laser frequencies are smaller than the repetition rate and can easily be spanned by an electro-optic modulator driven at MHz to GHz frequencies.\\

Such an RF-modulated pulse train might be used for laser cooling of carbon. In carbon, the wavelengths of the lowest dipole-allowed transitions lie blue of 170 nm, so one-photon cooling is no easier than for hydrogen. There are six states in the ground $2s^2 \: 2p^2$ electronic configuration, all having radiative lifetimes $> 1$ s and spanning an energy range of $12000 \:\:\mbox{cm}^{-1}$. The five singlet and triplet ground states remain decoupled from the quintet ground state under laser excitation, and one can avoid pumping into the $2s^2 2p^2 \:{}^1\mbox{S}_0$ state with a proper choice of cooling transitions. One-photon cooling thus would require four vacuum UV lasers, a formidable technical challenge. However, carbon has many two-photon transitions out of the ground-state manifold that can be excited with light in the 240 -- 270 nm range \cite{nistdata}, leading to the cooling cycle shown in Fig. 4. A single mode-locked laser can easily achieve the bandwidth needed for cooling on all four transitions. Second-order perturbation theory suggests transition rates of $10^{-3}$ to $10^{-5} \:\: I^2\:\: \mbox{Hz} \:\: \mbox{W}^{-2} \:\: \mbox{cm}^4$, orders of magnitude higher than for hydrogen 1S -- 2S, largely because of the relatively long upper-state lifetimes for carbon ($\sim 100$ ns). The cooling cycle of Fig. 2 involves excited states close to the ionization limit, for which excited-state photoionization can also be orders of magnitude smaller than for hydrogen \cite{opacity}. On the other hand, the recoil velocity and radiative lifetime both decrease an order of magnitude as compared to hydrogen. Because four transitions must be driven, the power available to drive each transition decreases a factor of four, while the necessity for four unequally spaced laser frequencies makes resonant enhancement of cooling power impractical. These advantages and disadvantages roughly balance for realistic parameter values, so laser cooling of carbon also appears feasible.\\

\begin{figure}
\begin{center}
\includegraphics*[width=\widtha]{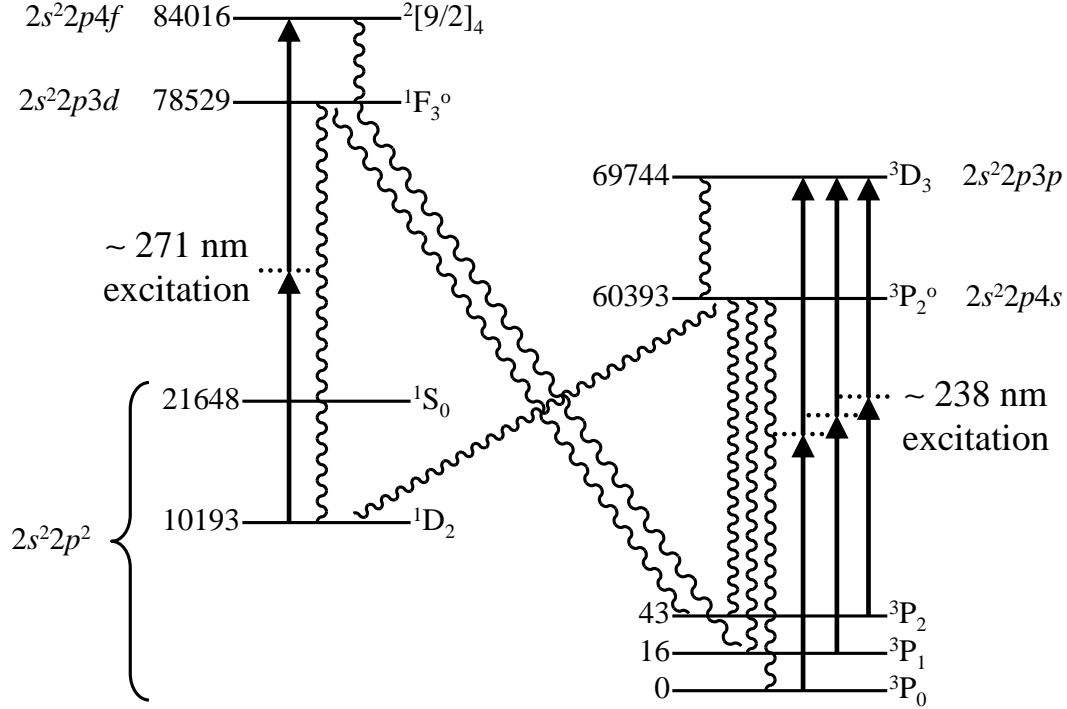}
\caption{Energy level diagram for laser cooling of carbon. Energies of states are given next to the horizontal line denoting the state, and are measured in $\mbox{cm}^{-1}$ above the lowest-energy state. Laser excitation is shown by solid vertical lines, radiative decay by wavy lines. Shaded boxes denote portions of the radiative decay paths that are not involved in laser excitation, and are labeled as belonging to the singlet or triplet manifold.}
\end{center}
\end{figure}

The cases of hydrogen and carbon suggest that mode-locked two-photon excitation can cool a variety of atomic species to temperatures $\sim 1$ mK if the atoms are precooled to a few hundred mK. Atomic and molecular gases have been cooled to these temperatures by thermalization with helium buffer gas \cite{bufferrev}. To obtain monatomic gases of refractory elements like carbon, one typically uses a hollow cathode discharge beam \cite{cbeam} which operates at high temperature. Buffer-gas cooling of such a beam, along the lines of \cite{bufferbeam}, provides a quite general precooling method for subsequent mode-locked two-photon cooling. In this case, new atomic species might be cooled to mK temperatures without the need for a complex and delicate superconducting magnetic trap.\\

Mode-locked two-photon excitation might also be useful in the laser cooling of trapped molecules, where it offers a route to ultracold temperatures without the loss of molecules associated with evaporative cooling. A buffer-gas magnetic trap has confined CaH at 400 mK \cite{Weinstein-Doyle-CaH-magnetic-trap}, while ND${}_3$ has been trapped in static electric fields at temperatures up to 300 mK \cite{Bethlem-Meijer-polar-molecule-trap,Rieger-Rempe-polar-molecule-trap}. Laser cooling a typical molecule requires exciting tens or hundreds of rovibrational levels, but some molecules have rovibrational structure that is relatively closed under repeated scattering. Single-photon laser cooling in CaH, for instance, might require as few as four cooling transitions \cite{DiRosa-laser-cooling-molecules}, and comparably closed two-photon cycles might also be identified for particular molecules. These cases seem amenable to cooling by RF-modulated pulse trains, as suggested above for carbon. More generally, as the number of metastable levels increases, the repetition rate of the laser must increase proportionately to keep all transitions resolved. Although cooling with a RF-modulated pulse train becomes ineffective in this case, Raman scattering in a molecular vapor can add sidebands to the cooling light that independently address the molecular rovibrational levels  \cite{msfcool}.\\

We have presented a new method of laser cooling based on two-photon excitation with ultrafast pulse trains. Pulse trains can provide the velocity selection necessary for laser cooling, and mode-locked light excites two-photon transitions as efficiently as CW laser light of the same average intensity. Frequency conversion is more efficient for ultrafast pulses, giving them an advantage for two-photon laser cooling of atoms whose lowest-energy single-photon transitions lie in the vacuum UV, such as H and \antih. It also seems possible to cool multielectron atoms, for instance carbon, by modulating a single pulse train at radio frequencies. In combination with buffer-gas precooling \cite{bufferrev,bufferbeam}, this method offers the chance to produce mK samples of a variety of new atomic species. The application of similar techniques to laser cooling of molecules is a tantalizing possibility.\\

\begin{acknowledgments}
The author would like to acknowledge the help afforded by discussions with Profs. D. Kleppner, F.X. K\"artner, and J.M. Doyle. The paper was greatly improved by insightful comments from E. Streed and Prof. D. Schneble. This work was supported by the US Air Force Office of Scientific Research under contract F49620-03-1-0313. The author was also supported by a Pappalardo Fellowship.
\end{acknowledgments}

\bibliography{bib1205}

\end{document}